\documentclass[a4paper,twoside,12pt]{article}
\input{obsjkt.sty}

\newcommand{\Msun}{\ensuremath{~{\rm M}_\odot}}                   % Solar mass symbol
\newcommand{\Rsun}{\ensuremath{~{\rm R}_\odot}}                   % Solar radius symbol
\newcommand{\rhosun}{\ensuremath{~\rho_\odot}}                    % Solar density symbol
\newcommand{\Teff}{\ensuremath{T_{\rm eff}}}                      % Effective temperature symbol
\newcommand{\logg}{\ensuremath{\log g}}                           % log(g) symbol
\newcommand{\Vsys}{\ensuremath{V_\gamma}}                         % Systemic velocity symbol
\newcommand{\EBV}{\ensuremath{E(B\!-\!V)}}                        % E(B-V) symbol
\newcommand{\Grp}{\ensuremath{G_{\rm RP}}}                        % Gaia G_RP band symbol
\newcommand{\degr}{\ensuremath{^\circ}}                           % Degree symbol
\renewcommand{\kms}{~km~s$^{-1}$}                                 % km/s symbol
\newcommand{\chisq}{\ensuremath{\chi^{\,2}}}                      % Chi-squared symbol
\newcommand{\chir}{\ensuremath{\chi_\nu^{\,2}}}                   % Reduced chi-squared symbol
\newcommand{\mc}[1]{\multicolumn{2}{c}{#1}}

\newcommand{\gaia}{\textit{Gaia}}
\newcommand{\targ}{ZZ~Boo}
\newcommand{\targfull}{ZZ Bo\"otis}

\newcommand{\Msunnom}{\hbox{$\mathcal{M}^{\rm N}_\odot$}}
\newcommand{\Rsunnom}{\hbox{$\mathcal{R}^{\rm N}_\odot$}}
\newcommand{\Lsunnom}{\hbox{$\mathcal{L}^{\rm N}_\odot$}}

\usepackage{rotating}

\begin{document} %%%%%%%%%%%%%%%%%%%%%%%%%%%%%%%%%%%%%%%%%%%%%%%%%%%%%%%%%%%%%%%%%%%%%%%%%%%%%%%%%%%%%%%%%%%%%%%%%%%%%%%%%%%%%%%%%%%%%%%%%%%%%%%%%%%%
%%%%%%%%%%%%%%%%%%%%%%%%%%%%%%%%%%%%%%%%%%%%%%%%%%%%%%%%%%%%%%%%%%%%%%%%%%%%%%%%%%%%%%%%%%%%%%%%%%%%%%%%%%%%%%%%%%%%%%%%%%%%%%%%%%%%%%%%%%%%%%%%%%%%%

\OBSheader{Rediscussion of eclipsing binaries: \targ}{J.\ Southworth}{2023 February}

\OBStitle{Rediscussion of eclipsing binaries. Paper XII. \\ The F-type twin system ZZ Bo\"otis}

\OBSauth{John Southworth}

\OBSinstone{Astrophysics Group, Keele University, Staffordshire, ST5 5BG, UK}

% \OBSabstract{Abs abs abs abs abs abs abs abs abs abs abs abs abs abs abs abs abs abs abs abs abs abs abs abs abs abs abs abs abs abs abs abs abs abs abs abs abs abs abs abs abs abs abs abs abs abs abs abs abs abs abs abs abs abs abs abs abs abs abs abs abs abs abs abs abs abs abs abs abs abs abs abs abs abs abs abs abs abs abs abs abs abs abs abs abs abs abs abs abs abs abs abs abs abs abs abs abs abs abs abs abs abs abs abs abs abs abs abs abs abs abs abs abs abs abs abs abs abs abs abs abs abs abs abs abs abs abs abs abs abs abs abs abs abs abs abs abs abs abs abs abs abs abs abs abs abs abs abs abs abs abs abs abs abs abs abs abs abs abs abs abs abs abs abs abs abs abs abs abs abs abs abs abs abs abs abs abs abs abs abs abs abs abs abs abs abs abs abs abs abs abs abs abs abs abs abs abs abs abs abs abs abs abs abs abs abs abs abs abs abs abs abs abs abs abs abs abs abs abs abs abs abs abs abs abs abs abs abs abs abs abs abs abs abs abs abs abs abs abs abs abs abs abs.}

\OBSabstract{\targ\ is an F-type detached eclipsing binary system containing two almost-identical stars on a circular orbit with a period of 4.992~d. We analyse light curves from two sectors of observations with the Transiting Exoplanet Survey Satellite (TESS) and two published sets of radial velocities of the component stars to determine their physical properties to high precision. We find masses of $1.558 \pm 0.008$\Msun\ and $1.599 \pm 0.012$\Msun, and radii of $2.063 \pm 0.006$\Rsun\ and $2.205 \pm 0.006$\Rsun. The similarity in the primary and secondary eclipse depths has led to confusion in the past. The high quality of the TESS\ data means we can, for the first time, clearly identify which is which. The primary star is conclusively hotter but smaller and less massive than the secondary star. We define a new high-precision orbital ephemeris and obtain effective temperatures using the \gaia\ parallax of the system. The secondary star is more evolved than the primary and a good agreement with theoretical predictions is found for a solar chemical composition and an age of 1.7~Gyr.}

%%%%%%%%%%%%%%%%%%%%%%%%%%%%%%%%%%%%%%%%%%%%%%%%%%%%%%%%%%%%%%%%%%%%%%%%%%%%%%%%%%%%%%%%%%%%%%%%%%%%%%%%%%%%%%%%%%%%%%%%%%%%%%%%%%%%%%%%%%%%%%%%%%%%%

\section*{Introduction}

Detached eclipsing binaries (dEBs) are a fundamental source of measured properties of normal stars \cite{Andersen91aarv,Torres++10aarv,Me15aspc} and are widely used to explore and calibrate our understanding of the properties of stars \cite{Andersen++90apj,ClaretTorres16aa,Tkachenko+20aa}. dEBs containing evolved stars are particularly helpful in tracing stellar evolution \cite{Andersen+88aa,Gallenne+16aa,Me21obs3}, especially if the two stars have similar masses but significantly different radii.

In this work we analyse a new space-based light curve and published radial velocities (RVs) of the dEB \targfull\ in order to determine its physical properties. The motivation for this series of papers is given in ref.~\cite{Me20obs}, and a review of the impact of space-based photometry can be found in ref.~\cite{Me21univ}.

\begin{table}[t]
\caption{\em Basic information on \targ. \label{tab:info}}
\centering
\begin{tabular}{lll}
{\em Property}                            & {\em Value}                 & {\em Reference}                   \\[3pt]
Right ascension (J2000)                   & 13:56:09.52                 & \cite{Gaia21aa}                   \\
Declination (J2000)                       & +25:55:07.4                 & \cite{Gaia21aa}                   \\
% Bright Star Catalogue                   & HR NNNN                     & \cite{HoffleitJaschek91}          \\
Henry Draper designation                  & HD 121648                   & \cite{CannonPickering20anhar}    \\
\textit{Hipparcos} designation            & HIP 68064                   & \cite{Hipparcos97}                \\
\textit{Tycho} designation                & TYC 2002-624-1              & \cite{Hog+00aa}                   \\
\textit{Gaia} DR3 designation             & 1450355965609917568         & \cite{Gaia21aa}                   \\
\textit{Gaia} DR3 parallax                & $9.3946 \pm 0.0324$ mas     & \cite{Gaia21aa}                   \\          % d = 106.44 +/- 0.36 pc
TESS\ Input Catalog designation          & TIC 357358259               & \cite{Stassun+19aj}               \\
$B$ magnitude                             & $7.158 \pm 0.015$           & \cite{Hog+00aa}                   \\          % \cite{Henden+15aas} for APASS
$V$ magnitude                             & $6.781 \pm 0.010$           & \cite{Hog+00aa}                   \\          % \cite{Hog+00aa} for Tycho
$J$ magnitude                             & $5.982 \pm 0.021$           & \cite{Cutri+03book}               \\
$H$ magnitude                             & $5.867 \pm 0.038$           & \cite{Cutri+03book}               \\
$K_s$ magnitude                           & $5.830 \pm 0.023$           & \cite{Cutri+03book}               \\
Spectral type                             & F3 V                        & \cite{Abt09apjs}                  \\[3pt]
\end{tabular}
\end{table}

\targ\ (Table~\ref{tab:info}) was found to be a spectroscopic binary by Shajn \cite{Shajn50izkry}, who presented the first RV curves of this object. Gaposchkin \citep{Gaposchkin51aj} announced the discovery of deep eclipses based on archival photographic plates. Gaposchkin \cite{Gaposchkin54aj} followed this up with a determination of the physical properties of the system based on a light curve from 1554 photographic plates and the RV curves from Shajn \cite{Shajn50izkry}. Miner \& McNamara \cite{MinerMcnamara63pasp} presented seven photographic spectra and derived orbits in agreement with those of Shajn \cite{Shajn50izkry}. McNamara et al.\ \cite{Mcnamara++71pasp} presented a photoelectric light curve with good coverage of the eclipses, which was subsequently reanalysed by Cester et al.\ \cite{Cester+78aas} and Botsula \cite{Botsula83pz}.

Popper \cite{Popper83aj} presented extensive spectroscopy of \targ\ based on photographic plates from the 3\,m Shane telescope at Lick Observatory. He analysed these, plus the McNamara et al.\ \cite{Mcnamara++71pasp} light curve, and determined the masses and radii of the system. Whilst the masses were very well established by the RVs, the radius measurements had errors of over 3\% due to the similarity of the stars and the limited quality of the photometry. Further RVs, which appear to be significantly more precise than those from Popper \cite{Popper83aj}, are advertised in a conference proceedings by Lacy \cite{Lacy92aspc} but remain unpublished. Finally, a good spectroscopic orbit has been presented by Nordstr\"om et al.\ \cite{Nordstrom+97aas}.

The spectral type of the system was given as F2\,V by Hill et al.\ \cite{Hill+75mmras} and as F3\,V by Abt \cite{Abt09apjs}. These supersede earlier assessments \cite{Shajn50izkry,MinerMcnamara63pasp}. Most recently, Kang et al.\ \cite{Kang+12aj} analysed a single high-resolution (resolving power $R \approx 80\,000$) \'echelle spectrum taken at orbital phase 0.583 to determine the effective temperatures (\Teff) and metallicities of the stars, and their light ratio. From a detailed chemical abundance analysis they determined that the primary star (star~A) has chemical abundances slightly lower than those of the Sun and suggested a similarity with the $\lambda$~Bo\"otis stars, although the abundance pattern is nowhere near extreme enough to match that class of chemically peculiar star \cite{Heiter02aa}. They found the secondary star (star~B) to have solar abundances with the exception of a $0.90 \pm 0.06$~dex overabundance of oxygen based on two spectral lines, and thus the two stars to have a significantly different abundance pattern.

In this work we revisit \targ\ to redetermine its masses and radii to high precision based on the plethora of published RVs and on a new space-based light curve. Detailed scientific motivations can be found in Refs.\ \cite{Me20obs} and \cite{Me21univ}.

% \textit{Discuss Vsini from popper, nordstrom, kang?}

%%%%%%%%%%%%%%%%%%%%%%%%%%%%%%%%%%%%%%%%%%%%%%%%%%%%%%%%%%%%%%%%%%%%%%%%%%%%%%%%%%%%%%%%%%%%%%%%%%%%%%%%%%%%%%%%%%%%%%%%%%%%%%%%%%%%%%%%%%%%%%%%%%%%%

\section*{Observational material}

\begin{figure}[t] \centering \includegraphics[width=\textwidth]{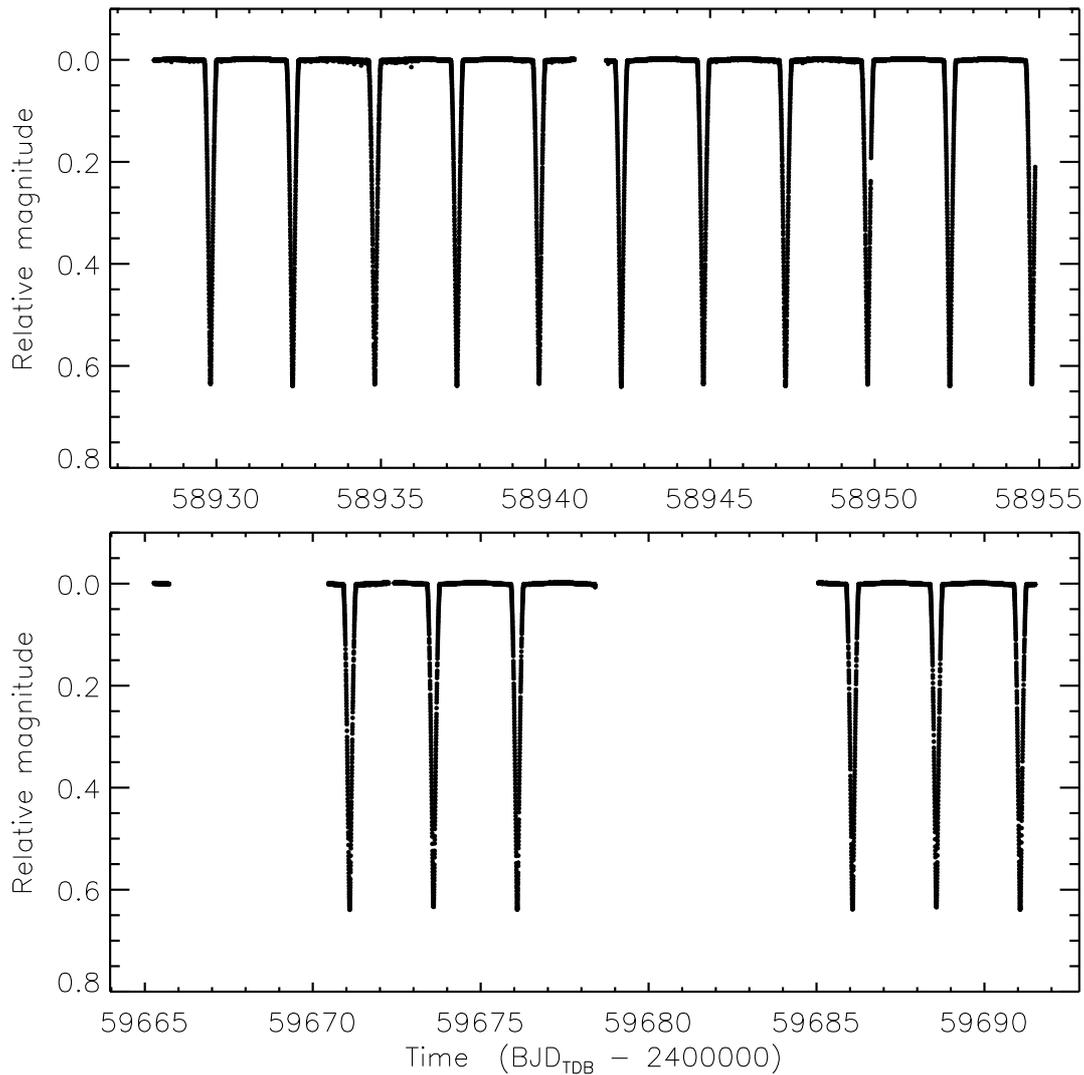} \\
\caption{\label{fig:time} TESS\ short-cadence SAP photometry of \targ\ from sectors 23
(top) and 50 (bottom). The flux measurements have been converted to magnitude units then
rectified to zero magnitude by the subtraction of low-order polynomials.} \end{figure}

% Sector 23 (2020-Mar-18 to 2020-Apr-16, in cycle 2): observed in camera 2.
% Sector 50 (2022-Mar-26 to 2022-Apr-22, in cycle 4): observed in camera 2.

The NASA Transiting Exoplanet Survey Satellite (TESS) observed \targ\ in sectors 23 (2020/03/18 to 2020/04/16) and 50 (2022/03/26 to 2022/04/22), in both cases in short cadence mode \cite{Ricker+15jatis}. The light curves show an uncomplicated light variation consisting of two eclipses of almost identical depth, plus a sinusoidal variation outside eclipse due to the ellipsoidal effect.

We downloaded these data from the MAST archive\footnote{Mikulski Archive for Space Telescopes, \\ \texttt{https://mast.stsci.edu/portal/Mashup/Clients/Mast/Portal.html}} and converted the fluxes to relative magnitude. The simple aperture photometry (SAP, ref.\ \cite{Jenkins+16spie}) is well-behaved and the pre-search data conditioning SAP (PDCSAP) data are errant, so we used only the SAP data in our analysis. We made no cut on the quality flag for sector 23 because even the data flagged as lower quality seemed to be of similar quality to the rest. Quite a lot of datapoints are not available for sector 50, and for this sector we required the QUALITY flag to be set to zero. Our analysis therefore included 18\,564 datapoints from sector 23 and 10\,456 from sector 50. We ignored the data errors as they are too small, preferring instead to determine the precision of the photometry from the scatter around the best-fitting model.

We queried the \gaia\ DR3 database\footnote{\texttt{https://vizier.cds.unistra.fr/viz-bin/VizieR-3?-source=I/355/gaiadr3}} in the region of \targ. A total of 17 additional sources are listed within 2~arcmin. The brightest of these is fainter than \targ\ by 7.78~mag in the \Grp\ passband (a light ratio of 0.00077) so we conclude that there is negligible contaminating light from nearby stars that are sufficiently distant from our target to be resolved by \gaia.

%%%%%%%%%%%%%%%%%%%%%%%%%%%%%%%%%%%%%%%%%%%%%%%%%%%%%%%%%%%%%%%%%%%%%%%%%%%%%%%%%%%%%%%%%%%%%%%%%%%%%%%%%%%%%%%%%%%%%%%%%%%%%%%%%%%%%%%%%%%%%%%%%%%%%

\begin{figure}[t] \centering \includegraphics[width=\textwidth]{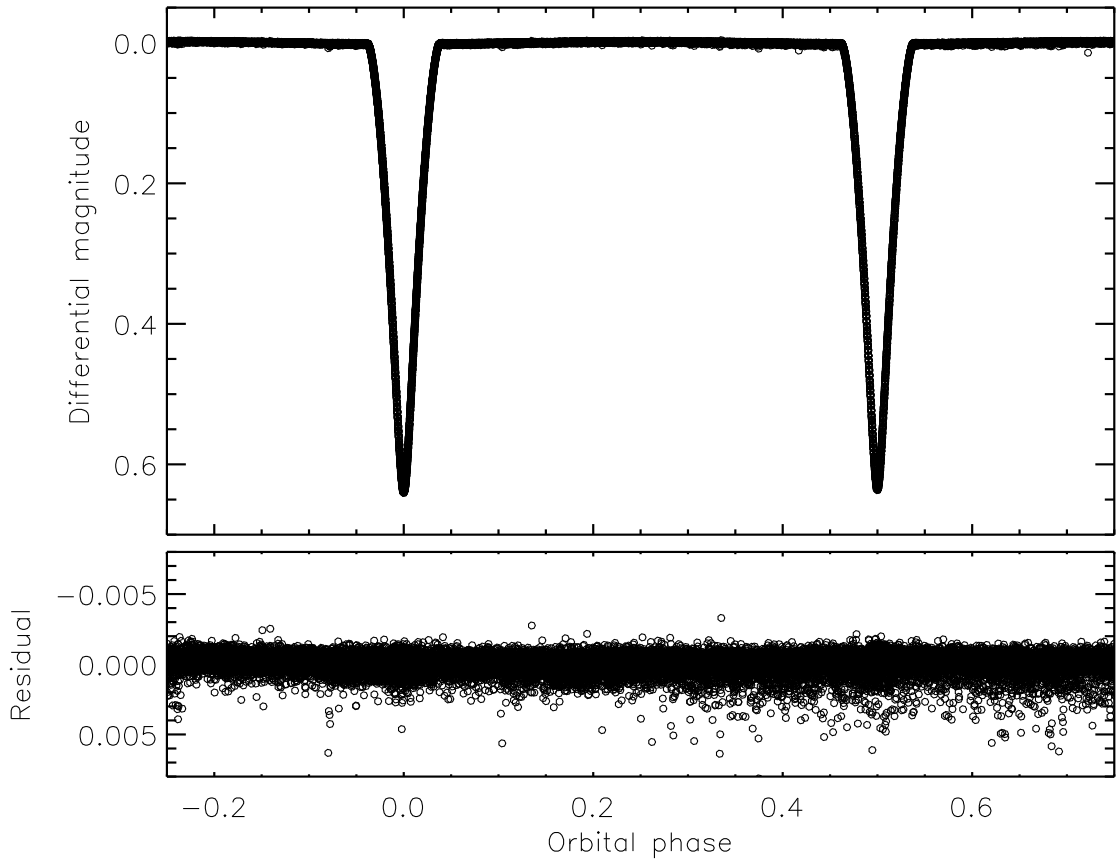} \\
\caption{\label{fig:phase} Best fit to the full TESS\ light curve of \targ\ using
{\sc jktebop}. The residuals are shown on an enlarged scale in the lower panel.} \end{figure}

\section*{Light curve analysis}

We first combined the SAP light curves from the two sectors. Then we removed three short stretches of data to avoid the possibility of them biasing the solution: the stretch between BJD 2458953.5 and 24594.9 because it contains an eclipse that is only partially covered; the small set of points around 2459665.5 because they are distant from other points and contain no eclipse; and the data in the interval 2459677.3 to 2459678.5 because they cover only out-of-eclipse phases and the first ten minutes of an otherwise-unobserved eclipse.

We designated the deeper of the two eclipses as the primary eclipse, at which time the primary star (hereafter star~A) is eclipsed by the secondary star (hereafter star~B). Although the eclipses are of very similar depth, the distinction between the two is clear in the TESS\ data. Based on this definition, star A is hotter, smaller and less massive than star~B. Past confusion as to which is the primary and secondary star is discussed below.

The TESS\ light curve of \targ\ was fitted using version 42 of the {\sc jktebop}\footnote{\texttt{http://www.astro.keele.ac.uk/jkt/codes/jktebop.html}} code \cite{Me++04mn2,Me13aa}. We fitted for the orbital period ($P$) and time of mid-eclipse ($T_0$), choosing as our reference time the primary eclipse closest to the midpoint of the data from sector 23. The fractional radii of the stars were included as their sum ($r_{\rm A}+r_{\rm B}$) and ratio ($k = {r_{\rm B}}/{r_{\rm A}}$), both of which were fitted, as were the orbital inclination ($i$) and the central surface brightness ratio of the two stars ($J$). After some tests we ruled out the presence of significant orbital eccentricity and thus adopted a circular orbit. Third light was found to be insignificant but was included as a fitted parameter to ensure its uncertainty was captured.

For limb darkening (LD) we adopted the quadratic law and forced the two stars to have the same coefficients due to their similarity. We fitted for the linear LD coefficient ($u_{\rm A,B}$) and fixed the quadratic LD coefficient ($v_{\rm A,B}$) to a theoretical value from Claret \cite{Claret17aa}.

\begin{table} \centering
\caption{\em \label{tab:jktebop} Adopted parameters of \targ\ measured from the
TESS\ light curves using the {\sc jktebop} code. The uncertainties are 1$\sigma$
and were determined using Monte Carlo and residual-permutation simulations.}
\begin{tabular}{lc}
{\em Parameter}                           &       {\em Value}                 \\[3pt]
{\it Fitted parameters:} \\
Time of primary eclipse (BJD$_{\rm TDB}$) & $ 2458942.300638 \pm  0.000004   $ \\
Orbital period (d)                        & $    4.99176522  \pm  0.00000010 $ \\
Orbital inclination (\degr)               & $      88.6361   \pm  0.0044     $ \\
Sum of the fractional radii               & $       0.23669  \pm  0.00008    $ \\
Ratio of the radii                        & $       1.0691   \pm  0.0014     $ \\
Central surface brightness ratio          & $       0.98003  \pm  0.00033    $ \\
Third light                               & $      -0.0001   \pm  0.0008     $ \\
Linear LD coefficient                     & $       0.246    \pm  0.005      $ \\
Quadratic LD coefficient                  &              0.22 (fixed)         \\
Orbital eccentricity                      &              0.0~ (fixed)         \\
{\it Derived parameters:} \\
Fractional radius of star~A               & $       0.11440  \pm  0.00011    $ \\
Fractional radius of star~B               & $       0.12230  \pm  0.00006    $ \\
Light ratio $\ell_{\rm B}/\ell_{\rm A}$   & $       1.1203   \pm  0.0029     $ \\[3pt]
\end{tabular}
\end{table}

% IDL> [0.00011,0.00006]/[0.114397,0.122297]*100
%      0.096156374     0.049060892

The best fit is shown in Fig.~\ref{fig:phase} and is extremely good. The residuals show a slight excess to fainter magnitudes: these datapoints are a subset of those flagged as less reliable in sector 23. Upon investigation we found that they have a negligible effect on the solution, so we did not reject the flagged data. The fitted parameters are given in Table~\ref{tab:jktebop}. %The light ratio is in excellent agreement with that found spectroscopically by Kang et al.\ \cite{Kang+12aj}.

%%%%%%%%%%%%%%%%%%%%%%%%%%%%%%%%%%%%%%%%%%%%%%%%%%%%%%%%%%%%%%%%%%%%%%%%%%%%%%%%%%%%%%%%%%%%%%%%%%%%%%%%%%%%%%%%%%%%%%%%%%%%%%%%%%%%%%%%%%%%%%%%%%%%%

\section*{Uncertainties in the photometric parameters}

\begin{figure}[t] \centering \includegraphics[width=\textwidth]{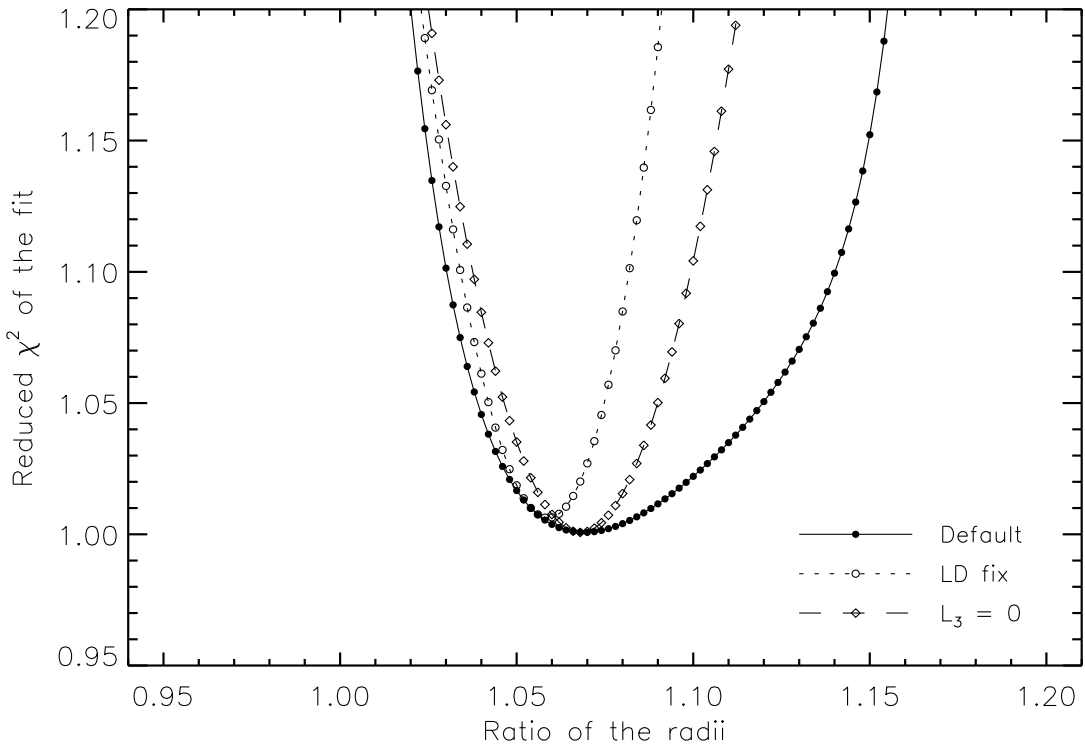} \\
\caption{\label{fig:kfix} Quality of the fit to the TESS\ data of \targ\ as a
function of the ratio of the radii. Three different sets of solutions are plotted,
as given in the legend.} \end{figure}

To determine the uncertainties of the fitted parameters we ran 10\,000 Monte Carlo and residual-permutation simulations \cite{Me++04mn2,Me08mn} using {\sc jktebop} tasks 8 and 9. In past work we have found that these uncertainty estimates are reliable \cite{Maxted+20mn,Me21obs4,Me21obs5}. The residual-permutation simulations return slightly larger errorbars, possibly due to the non-Gaussian nature of the residuals, so were used as the final uncertainties (Table~\ref{tab:jktebop}).

The uncertainties are extremely small, and beyond the level of precision to which we consider the light curve model reliable. Maxted et al.\ \cite{Maxted+20mn} demonstrated a precision of 0.2\% in the radii of the EB AI~Phe, which is totally-eclipsing so is better-suited to such measurements. Based on this, we recommend imposing a minimum uncertainty of 0.2\% on $r_{\rm A}$ and $r_{\rm B}$.

Due to the small values of the uncertainties, we explored whether the solution really is as well-determined as it seems. We ran a set of fits in the same way as above but with $k$ fixed at values between 0.90 and 1.20 in steps of 0.002. We chose this parameter as it is correlated with all other parameters of interest in the {\sc jktebop} fit ($r_{\rm A}$, $r_{\rm B}$, $i$, $J$, $u_{\rm A,B}$). We assigned a single errorbar of size 0.702~mmag to every TESS\ datapoint to give a reduced \chisq\ of $\chir = 1.0$ for the best fit found above. A plot of the results (Fig.~\ref{fig:kfix}) shows that there is a single well-defined \chir\ minimum at the best-fitting $k$ found above.

We then ran a set of fits for the same grid of $k$ values but with the LD fixed to the theoretically-predicted values of $u_{\rm A,B} = 0.30$ and $v_{\rm A,B} = 0.22$. This is also plotted in Fig.~\ref{fig:kfix} and shows a much narrower minimum in \chir\ at a slightly lower value of $k$. The best fit with fixed LD coefficients has an $r_{\rm A}$ lower by 0.81\%, an $r_{\rm B}$ higher by 0.14\%, and a \chisq\ larger by 151. We conclude that this additional dependence on stellar theory produces a slightly different and worse fit that can be safely ignored, and that it is better to fit for LD when it is well determined by the available data.

We also ran a set of fits that were the same as the default solution but assuming $L_3 = 0$, which are also shown in Fig.~\ref{fig:kfix}. The overall best fit is almost identical to the default solution, but the \chir\ minimum is narrower because the fit is more constrained. Based on these tests, we are confident that the parameters in Table~\ref{tab:jktebop} are reliable and that the measured uncertainties are not clearly underestimated (apart from our argument above for a minimum of 0.2\% on the fractional radii).

\section*{Orbital ephemeris}

\begin{figure}[t] \centering \includegraphics[width=\textwidth]{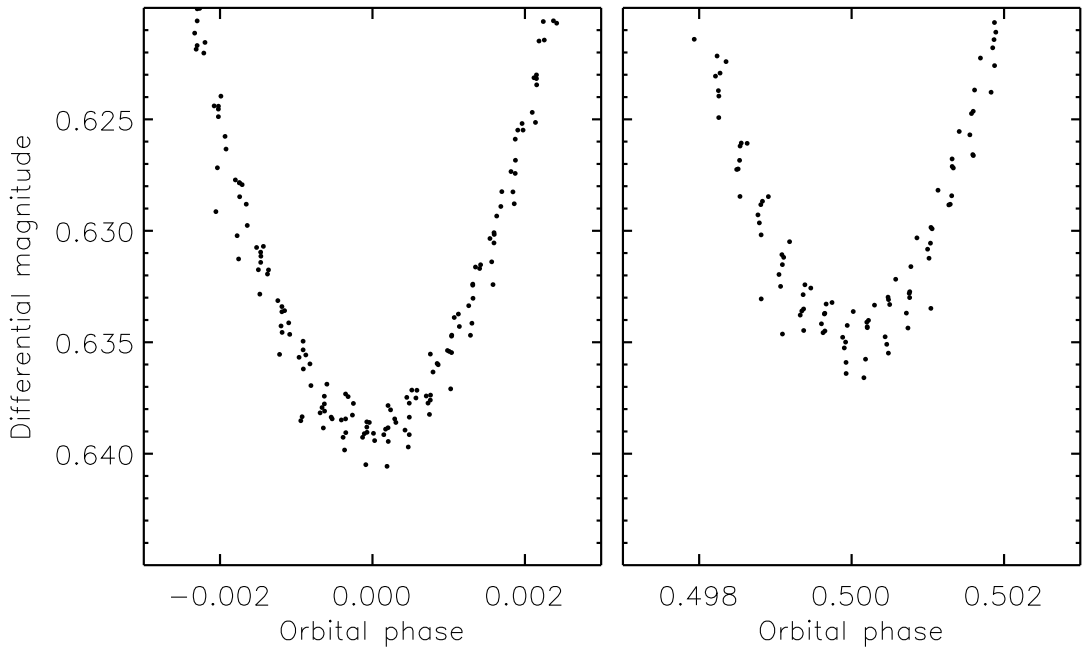} \\
\caption{\label{fig:ecl} Extreme close-up of the primary (left) and secondary (right)
eclipses in the TESS\ data to show their different depths.} \end{figure}

We now had a precise set of photometric parameters and a decent orbital ephemeris. However, a higher-precision ephemeris defined over a longer timescale would be useful in our analysis of the RVs in the following section. We therefore measured the times of individual eclipses in the TESS\ data and performed a literature search to obtain reliable eclipse times for earlier epochs. All published epochs were either stated or assumed to be on the HJD$_{\rm UTC}$ timescale and thus converted to BJD$_{\rm TDB}$ before analysis.

\begin{table} \centering
\caption{\em Times of published mid-eclipse for \targ\ and their residuals versus the fitted ephemeris.\label{tab:tmin}}
\setlength{\tabcolsep}{10pt}
\begin{tabular}{rr@{.}lr@{.}lr@{.}ll}
{\em Orbital} & \mc{\em Eclipse time} & \mc{\em Uncertainty} & \mc{\em Residual} & {\em Reference} \\
{\em cycle}   & \mc{\em (BJD$_{TDB}$)}    & \mc{\em (d)}          & \mc{\em (d)}     &              \\[3pt]
$-4082.0$ & 2438565&9196   & ~~~~~0&0100   &   $ $0&0045   & \cite{Mcnamara++71pasp} \\   %   390.3 s    0.3 sigma
$-3208.0$ & 2442928&7177   &      0&0004   & ~~$-$0&0002   & \cite{Samolyk09javso}   \\   %   -20.6 s   -0.4 sigma
$-1237.0$ & 2452767&4840   &      0&0016   &   $-$0&0031   & \cite{Zejda04ibvs}      \\   %  -264.0 s   -1.3 sigma
$ -800.5$ & 2454946&39396  &      0&00050  &   $ $0&00136  & \cite{Brat+09oejv}      \\   %   117.6 s    1.8 sigma
$ -720.5$ & 2455345&7334   &      0&0007   &   $-$0&0004   & \cite{Samolyk11javso}   \\   %   -37.6 s   -0.4 sigma
$   -2.5$ & 2458929&821188 &      0&000010 &   $-$0&000040 & This work               \\   %    -3.5 s   -2.7 sigma
$   -2.0$ & 2458932&317080 &      0&000008 &   $-$0&000031 & This work               \\   %    -2.7 s   -2.6 sigma
$   -1.5$ & 2458934&813016 &      0&000019 &   $ $0&000022 & This work               \\   %     1.9 s    0.8 sigma
$   -1.0$ & 2458937&308911 &      0&000007 &   $ $0&000035 & This work               \\   %     3.0 s    3.3 sigma
$   -0.5$ & 2458939&804750 &      0&000010 &   $-$0&000009 & This work               \\   %    -0.8 s   -0.6 sigma
$    0.0$ & 2458942&300615 &      0&000007 &   $-$0&000026 & This work               \\   %    -2.3 s   -2.5 sigma
$    0.5$ & 2458944&796554 &      0&000009 &   $ $0&000030 & This work               \\   %     2.6 s    2.2 sigma
$    1.0$ & 2458947&292415 &      0&000012 &   $ $0&000008 & This work               \\   %     0.7 s    0.5 sigma
$    1.5$ & 2458949&788231 &      0&000014 &   $-$0&000058 & This work               \\   %    -5.0 s   -2.8 sigma
$    2.0$ & 2458952&284195 &      0&000006 &   $ $0&000023 & This work               \\   %     2.0 s    2.6 sigma
$  146.0$ & 2459671&098352 &      0&000010 &   $-$0&000008 & This work               \\   %    -0.7 s   -0.5 sigma
$  146.5$ & 2459673&594232 &      0&000005 &   $-$0&000011 & This work               \\   %    -0.9 s   -1.4 sigma
$  147.0$ & 2459676&090125 &      0&000005 &   $-$0&000000 & This work               \\   %    -0.0 s   -0.0 sigma
$  149.0$ & 2459686&073665 &      0&000006 &   $ $0&000009 & This work               \\   %     0.8 s    1.1 sigma
$  149.5$ & 2459688&569536 &      0&000005 &   $-$0&000002 & This work               \\   %    -0.2 s   -0.3 sigma
$  150.0$ & 2459691&065431 &      0&000006 &   $ $0&000010 & This work               \\   %     0.9 s    1.1 sigma
\end{tabular}
\end{table}

We paid careful attention to ensuring that we chose the deeper of the two types of eclipse as the primary. The high quality of the TESS\ data makes this choice definitive, for the first time, as the eclipse depths can be determined to be 0.639~mag for the primary eclipse and 0.634~mag for the secondary eclipse (see Fig.~\ref{fig:ecl}). The similarity of these numbers has led to confusion in the past (see discussions in refs.\ \cite{Popper83aj} and \cite{Cester+78aas}) and demands care in interpreting orbital phases in past publications.

Once we had assembled the available times of minimum we used the ephemeris from only the TESS\ data in Table~\ref{tab:jktebop} to assign cycle numbers and eclipse types (primary or secondary) to them. We then fit them with a straight line to obtain a final orbital ephemeris:
\begin{equation}
  \mbox{Min~I} = {\rm BJD}_{\rm TDB}~ 2458942.300641 (7) + 4.991765196 (61) E
\end{equation}
which is very precise because the eclipses are deep and V-shaped so are excellent fiducials, and because the first and last eclipses in the TESS\ data are separated by 761~d. The times of minimum and their residual versus the final ephemeris are given  in Table~\ref{tab:tmin}. There is no evidence for nonlinearity in the eclipse timings, in the sense that a quadratic fit to the timings gives an almost identical fit with a quadratic coefficient much smaller than its uncertainty.

We have extrapolated this ephemeris back to times of eclipse given by past authors to see how it compares to previous work. The time of primary minimum given by Popper \cite{Popper83aj} actually corresponds to a secondary minimum, so the masses of the two stars quoted by him should be swapped. The times of primary eclipse given by Miner \& McNamara \cite{MinerMcnamara63pasp} and McNamara et al.\ \cite{Mcnamara++71pasp} are indeed primary eclipses according to our ephemeris.

%%%%%%%%%%%%%%%%%%%%%%%%%%%%%%%%%%%%%%%%%%%%%%%%%%%%%%%%%%%%%%%%%%%%%%%%%%%%%%%%%%%%%%%%%%%%%%%%%%%%%%%%%%%%%%%%%%%%%%%%%%%%%%%%%%%%%%%%%%%%%%%%%%%%%

\section*{Radial velocities}

\begin{figure}[t] \centering \includegraphics[width=\textwidth]{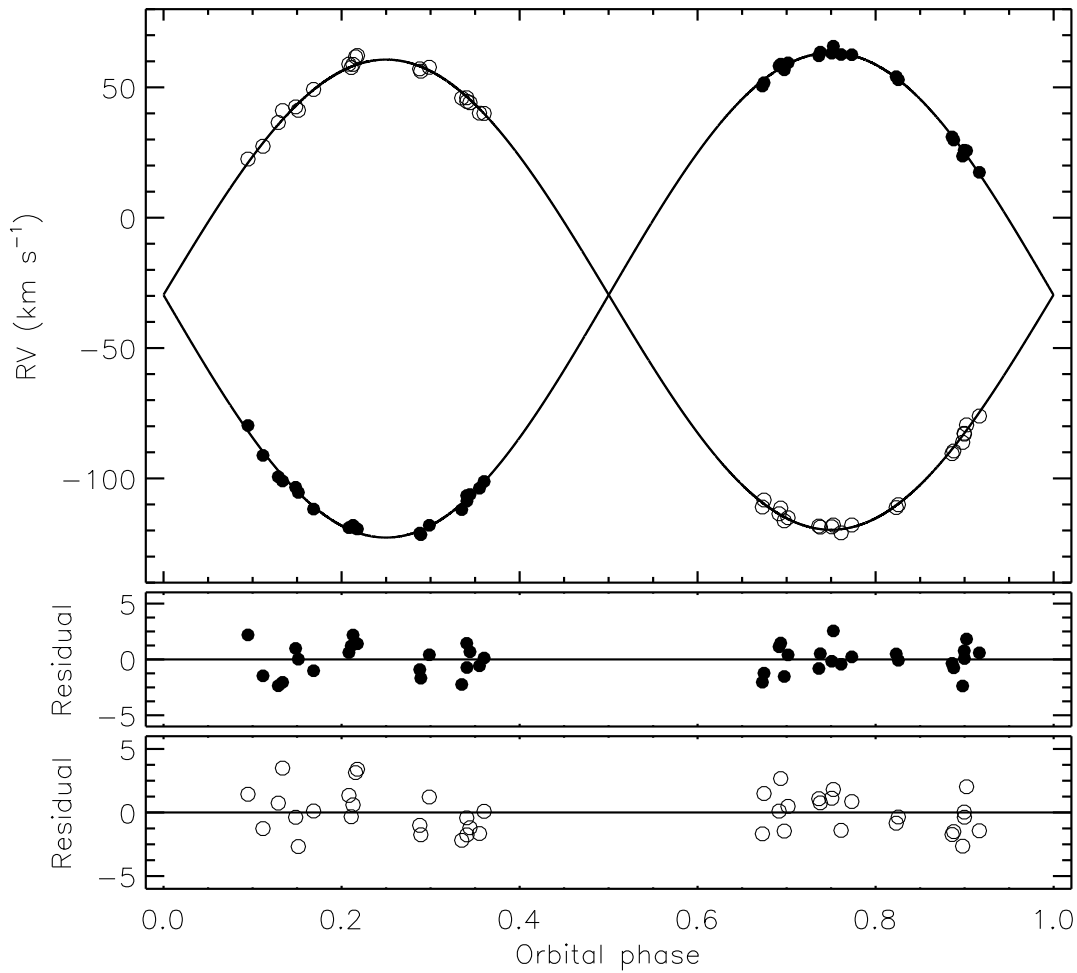} \\
\caption{\label{fig:popper} RVs of \targ\ measured by Popper \cite{Popper83aj}
(filled circles for star~A and open circles for star~B) compared to the best-fitting
spectroscopic orbits from {\sc jktebop} (solid curves). The residuals are given in
the lower panels separately for the two components.} \end{figure}

\begin{figure}[t] \centering \includegraphics[width=\textwidth]{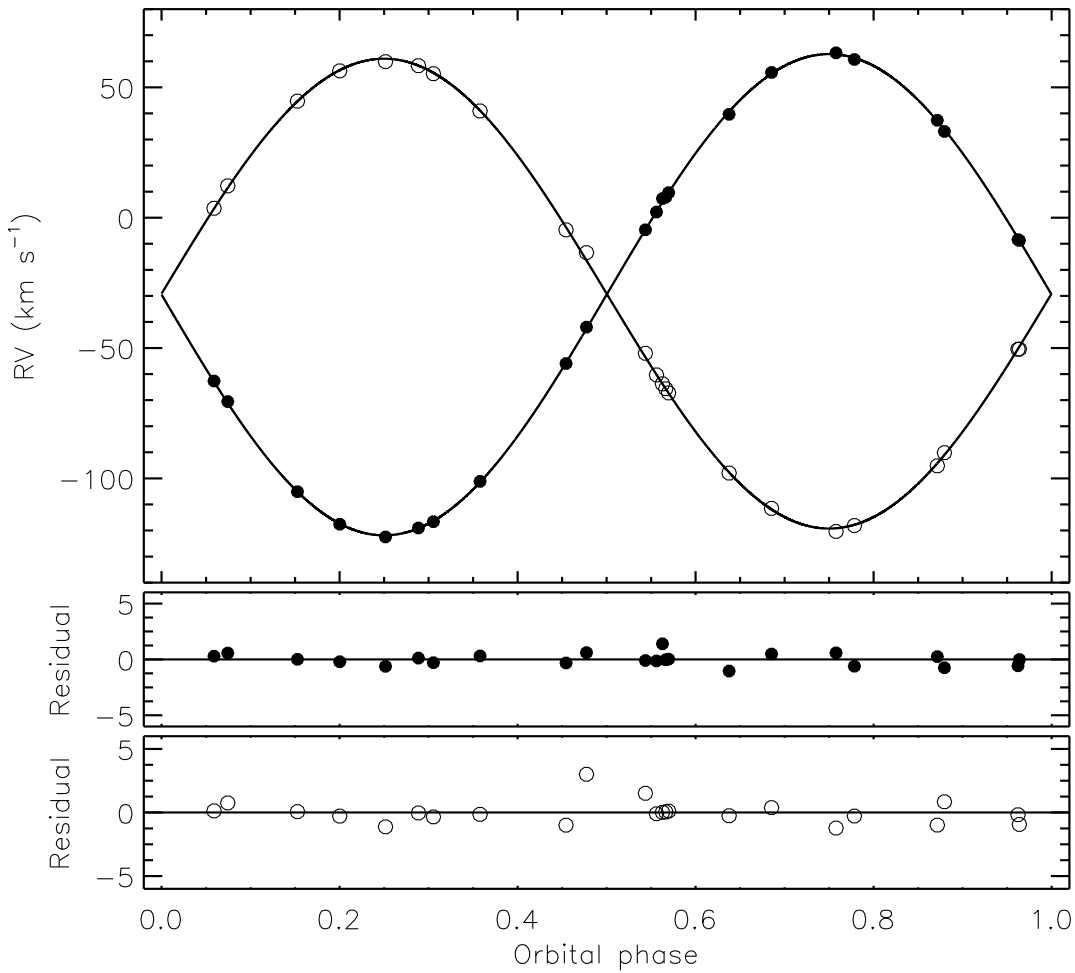} \\
\caption{\label{fig:nordstrom} RVs of \targ\ measured by Nordstr\"om et al.\
\cite{Nordstrom+97aas}. Other comments are as in Fig.~\ref{fig:popper}.} \end{figure}

Three sets of RVs have been published for \targ. Those from Popper \cite{Popper83aj} comprise 42 per star, were tabulated in that work, and were re-analysed here. Those from Nordstr\"om et al.\ \cite{Nordstrom+97aas} include 30 RVs per star, are available electronically from the CDS, and were also re-fitted here. Those from Lacy \cite{Lacy92aspc} are available only in plot form; we have been unable to access the original measurements so did not use them in the current work.

The Popper \cite{Popper83aj} RVs were fitted using {\sc jktebop} using the orbital period from the previous section and fitting for the velocity amplitude and systemic velocity of each star individually. We also fitted for the ephemeris zeropoint to allow for any inaccuracies in the ephemeris or reported timestamps, in light of our experience with ZZ\,UMa \cite{Me22obs5}. The Popper RVs are tabulated with time to three decimal places, RV to one decimal place, and no uncertainties. We therefore weighted the RVs for individual stars equally. We found, as expected, that we had to swap the identity of the two stars due to the different choice of which is the primary star. Uncertainties in the fitted parameters were determined using Monte Carlo simulations (see Paper\,VI, ref.\cite{Me21obs5}). Our results are given in Table~\ref{tab:orbits} and are in excellent agreement with those of Popper \cite{Popper83aj}. Our systemic velocities are slightly lower: this is due to the addition of a $+1.4$\kms\ correction by Popper from observations of an RV standard star, which he applied to his systemic velocities but not the individual RVs. The RVs and the best fits are shown in Fig~\ref{fig:popper}.

For the Nordstr\"om et al.\ \cite{Nordstrom+97aas} RVs we proceeded in the same way but were forced to reject some observations. A subset of the spectra were obtained during eclipse and show a much larger scatter around the best fit, so we rejected all spectra taken within $\pm$0.03 orbital phases of the midpoint of an eclipse. We also rejected the RVs from epoch 2447922.9251 because both were 5\kms\ closer to the systemic velocity than predicted by the best fit -- this could have been caused by an incorrect timestamp or a problem with the observation itself. This left 23 RVs per star, which were fitted as were the Popper RVs. We had to swap the identities of the two stars here as well. The results are given in Table~\ref{tab:orbits}. Our results are in reasonable agreement with those of Nordstr\"om et al.\ \cite{Nordstrom+97aas}, but with some differences due to our rejection of data we considered unhelpful. The RVs and the best fits are shown in Fig.~\ref{fig:nordstrom}.

Table~\ref{tab:orbits} shows that the spectroscopic orbits from the two sets of RVs agree well for star~B but not for star~A. The disagreement is lower than if we had adopted all of the Nordstr\"om RVs rather than rejecting those we considered to be less reliable. We decided that the best option was to combine the two velocity amplitudes for each star via a weighted mean. We then multiplied the errorbar in the velocity amplitude of star~A by the square-root of the \chir\ of the average to account for the small discrepancy between the two datasets. The results remain high-quality measurements of the orbital motion of the two stars.

\begin{table} \centering
\caption{\em \label{tab:orbits} Spectroscopic orbits for \targ\ from the literature
and from the reanalysis of the RVs in the current work. All quantities are in\kms.
The values from Popper \cite{Popper83aj} and Nordstr\"om et al.\ \cite{Nordstrom+97aas}
have each been swapped to account for their different identification of which is the primary star.}
\begin{tabular}{lrrrrrc}
{\em Source}  & $K_{\rm A}$~ & $K_{\rm B}$~ & ${\Vsys}$~ & ${\Vsys}_{\rm ,A}$~ & ${\Vsys}_{\rm ,B}$~ & $rms$ residual \\[7pt]

Popper \cite{Popper83aj}                  &     93.1  &     90.2  &           &  $-$28.4  &  $-$28.1  &            \\
                                          & $\pm$0.2  & $\pm$0.3  &           & $\pm$0.2  & $\pm$0.3  &            \\
This work                                 &     92.98 &     90.18 &           &  $-$29.72 &  $-$29.53 & 1.30, 1.60 \\
                                          & $\pm$0.22 & $\pm$0.31 &           & $\pm$0.21 & $\pm$0.25 &            \\[7pt]
Nordstr\"om et al.\ \cite{Nordstrom+97aas} &     92.02 &     89.85 &  $-$29.50 &           &           & 1.86, 2.39 \\
                                          & $\pm$0.72 & $\pm$0.56 & $\pm$0.27 &           &           &            \\
This work                                 &     92.31 &     90.12 &           &  $-$29.55 &  $-$29.12 & 0.53, 0.90 \\
                                          & $\pm$0.16 & $\pm$0.29 &           & $\pm$0.11 & $\pm$0.19 &            \\[7pt]
Final values                              &     92.54 &     90.14 &           &           &           &            \\
                                          & $\pm$0.32 & $\pm$0.17 &           &           &           &            \\
\end{tabular}
\end{table}

% print, wmean([92.98,92.31],[0.22,0.16],chisq=chisq,sigma=sigma),chisq,sigma,sigma*sqrt(chisq)
%       92.5418      6.06632     0.129398     0.318705
% print, wmean([90.18,90.12],[0.31,0.20],chisq=chisq,sigma=sigma),chisq,sigma,sigma*sqrt(chisq)
%       90.1376    0.0264490     0.168059    0.0273317

%%%%%%%%%%%%%%%%%%%%%%%%%%%%%%%%%%%%%%%%%%%%%%%%%%%%%%%%%%%%%%%%%%%%%%%%%%%%%%%%%%%%%%%%%%%%%%%%%%%%%%%%%%%%%%%%%%%%%%%%%%%%%%%%%%%%%%%%%%%%%%%%%%%%%

\section*{Chromospheric emission}

\begin{figure}[t] \centering \includegraphics[width=\textwidth]{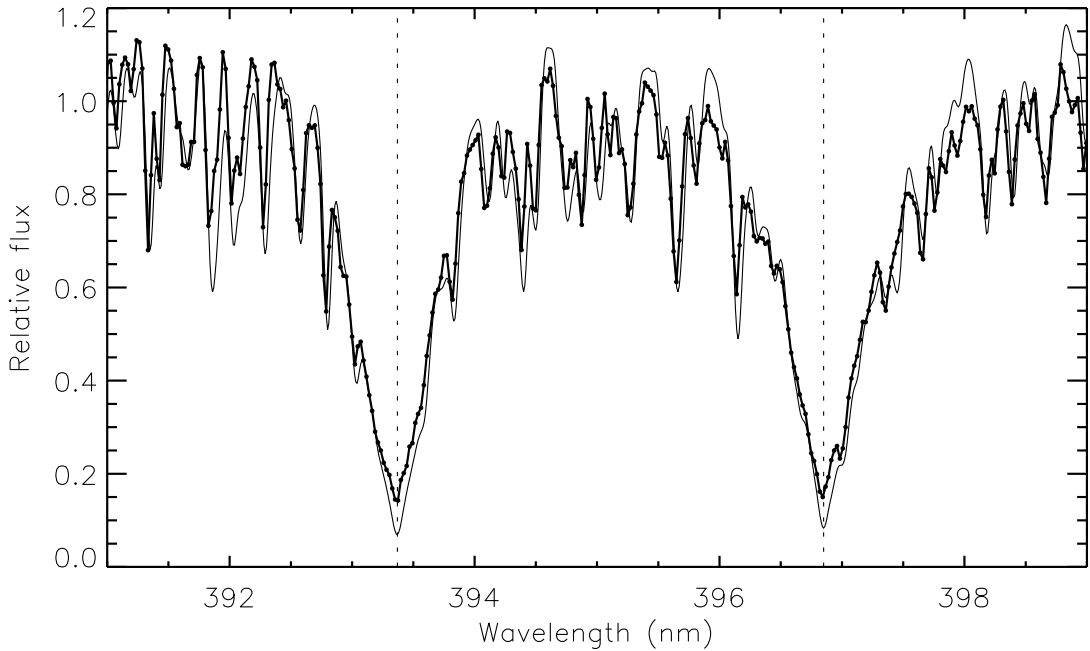} \\
\caption{\label{fig:cahk} Observed spectrum of \targ\ around the Ca~{\sc ii} H and K lines (thick upper
line with points) compared to a synthetic spectrum for a star with $\Teff = 6700$~K, $\logg = 4.0$ and
solar metallicity from the BT-Settl model atmospheres \cite{Allard+01apj,Allard++12rspta} (thin line
without points). The H and K line central wavelengths are shown with dotted lines. The spectrum of
\targ\ has been shifted to zero velocity and normalised to unit flux.} \end{figure}

In order to investigate the possibility of magnetic activity, the Ca~{\sc ii} H and K lines of several dEBs in this series have been observed using the Intermediate Dispersion Spectrograph (IDS) at the Cassegrain focus of the Isaac Newton Telescope (INT); see Paper\,XI, ref.~\cite{Me22obs5}. \targ\ is not a promising target for chromospheric emission due to its relatively high \Teff, but was nevertheless included as its brightness meant a good spectrum could be obtained using minimal observing time. A single observation of 120~s duration was obtained on the night of 2022/06/07 in excellent weather conditions. We used the 235~mm camera, H2400B grating, EEV10 CCD and a 1~arcsec slit and obtained a resolution of approximately 0.05~nm. A central wavelength of 4050\,\AA\ yielded a spectrum covering 373--438~nm at a reciprocal dispersion of 0.023~nm~px$^{-1}$. The data were reduced using a pipeline currently being written by the author, which performs bias subtraction, division by a flat-field from a tungsten lamp, aperture extraction, and wavelength calibration using copper-argon and copper-neon arc lamp spectra.

The spectrum was obtained at orbital phase 0.5101 and is shown in Fig.~\ref{fig:cahk}. The Ca H and K line centres exhibit a higher flux than the synthetic spectrum provided for comparison, but this can be attributed to the rotational velocities of the component stars plus the velocity difference of 11.2\kms\ between them at the time of the observation. There is no clear evidence for chromospheric emission (as expected) or for spot activity in the light curves. We conclude that \targ\ does not show magnetic activity detectable with the currently available data.

% print, (59738.53746d0-58942.300641d0)/4.991765196d0 mod 1.0d0
%       0.51007063
% 0.51010  0.449829   0.50598  0.49096  0.00000  0.449829  -23.6930  -34.8435
% print, 34.8435-23.6930
%       11.1505

%%%%%%%%%%%%%%%%%%%%%%%%%%%%%%%%%%%%%%%%%%%%%%%%%%%%%%%%%%%%%%%%%%%%%%%%%%%%%%%%%%%%%%%%%%%%%%%%%%%%%%%%%%%%%%%%%%%%%%%%%%%%%%%%%%%%%%%%%%%%%%%%%%%%%

\section*{Physical properties of \targ}

% Central surface brightness ratio          & $       0.98003  \pm  0.00033    $ \\
% Teff ratio: 0.994970 pm 0.000084
% Teff diff: if Teff1=6700 then Teff2=6666.3pm 0.6

We calculated the physical properties of \targ\ using quantities determined from the light curve (Table~\ref{tab:jktebop}) and RVs (Table~\ref{tab:orbits}), standard formulae \cite{Hilditch01book} and the reference solar values from the IAU \cite{Prsa+16aj}. The errorbars on $r_{\rm A}$ and $r_{\rm B}$ were increased to 0.2\% following the discussion above. We used the {\sc jktabsdim} code \cite{Me++05aa}, which propagates uncertainties using a perturbation approach. The results are given in Table~\ref{tab:absdim} and show that the masses are determined to 0.5\% and 0.8\%, and the radii to 0.2\% precision. The relatively low precision of $K_{\rm A}$, due to the minor disagreement between the two sources of published RVs, is the main source of uncertainty in the masses. The radii are measured to a precision an order of magnitude better than the previous determination by Popper \cite{Popper83aj}, due to the high quality of the TESS\ light curve. The measurements of \targ\ are now good enough for it to be included in the Detached Eclipsing Binary Catalogue (DEBCat\footnote{\texttt{https://www.astro.keele.ac.uk/jkt/debcat/}}, ref. \cite{Me15aspc})

The \Teff\ values of the stars are very similar: from the surface brightness ratio we find a \Teff\ ratio of $0.99497 \pm 0.00008$. The F3\,V spectral type of the system corresponds to a \Teff\ of 6720~K (ref.\cite{PecautMamajek13apjs}),  Popper \cite{Popper83aj} gave $6669 \pm 31$~K for both stars, and Nordstr\"om et al.\ \cite{Nordstrom+97aas} used template spectra at 6750~K for their RV measurements. However, rather higher \Teff\ measurements of $6860 \pm 20$~K and $6930 \pm 20$~K were found by Kang et al.\ \cite{Kang+12aj}, prompting us to obtain our own values. We thus determined the distance to the system using the surface brightness versus \Teff\ relations from Kervella et al.\ \cite{Kervella+04aa}, the $K$-band apparent magnitude from 2MASS (Table~\ref{tab:info}), and adopting an interstellar extinction of $\EBV = 0.00 \pm 0.01$ because the system is close to the Sun. We found the best agreement with the \gaia\ parallax distance ($106.44 \pm 0.36$~pc) if the stars have a \Teff\ of 6705~K, and assigned a conservative uncertainty of 100~K. Accounting for the small \Teff\ difference we therefore adopted values of 6720 and 6690~K. An improved \Teff\ measurement using the \gaia\ parallax and apparent magnitudes will be presented in future.

\begin{table} \centering
\caption{\em Physical properties of \targ\ defined using the nominal solar units given by IAU
2015 Resolution B3 (ref.\ \cite{Prsa+16aj}). \label{tab:absdim}}
\begin{tabular}{lr@{\,$\pm$\,}lr@{\,$\pm$\,}l}
{\em Parameter}        & \multicolumn{2}{c}{\em Star A} & \multicolumn{2}{c}{\em Star B}    \\[3pt]
Mass ratio                                  & \multicolumn{4}{c}{$1.0266 \pm 0.0040$}       \\
Semimajor axis of relative orbit (\Rsunnom) & \multicolumn{4}{c}{$18.024 \pm 0.035$}        \\
Mass (\Msunnom)                             &  1.5572 & 0.0080      &  1.599  & 0.012       \\
Radius (\Rsunnom)                           &  2.0626 & 0.0057      &  2.2050 & 0.0062      \\
Surface gravity ($\log$[cgs])               &  4.0016 & 0.0019      &  3.9550 & 0.0023      \\
Density ($\!\!$\rhosun)                     &  0.1775 & 0.0011      &  0.1491 & 0.0009      \\
Synchronous rotational velocity ($\!\!$\kms)& 20.905  & 0.058       &  22.348 & 0.062       \\
Effective temperature (K)                   &   6720  & 100         &   6690  & 100         \\
Luminosity $\log(L/\Lsunnom)$               &   0.893 & 0.026       &   0.943 & 0.026       \\
$M_{\rm bol}$ (mag)                         &   2.507 & 0.065       &   2.382 & 0.066       \\
Distance (pc)                               & \multicolumn{4}{c}{$106.5 \pm 1.4$}           \\[3pt]
\end{tabular}
\end{table}

% print, [0.0080,0.012,0.0045,0.0045]/[1.5572,1.599,2.0626,2.2050]*100
%      0.513743     0.750469     0.218171     0.204082

\begin{figure}[t] \centering \includegraphics[width=\textwidth]{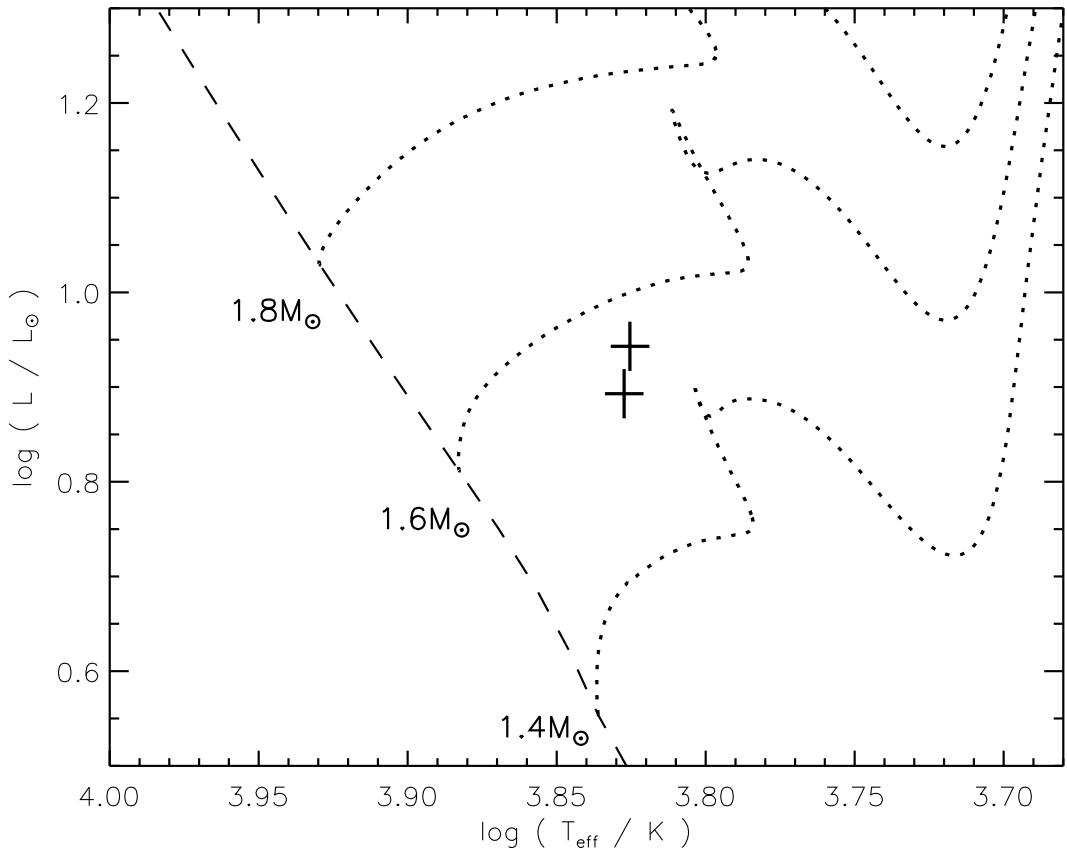} \\
\caption{\label{fig:hrd} Hertzsprung-Russell diagram showing the components of \targ\
(solid crosses) and selected predictions from the PARSEC models \cite{Bressan+12mn}
(dotted lines) beginning at the zero-age main sequence (dashed line). Models for 1.4, 1.6
and 1.8\Msun\ are shown (labelled), for a metal abundance of $Z=0.017$.} \end{figure}

From Table~\ref{tab:absdim} we can see that the primary star is hotter, but smaller and less massive than its companion. The higher \Teff\ is confirmed to very high significance from the surface brightness ratio in Table~\ref{tab:jktebop}. Both are significantly evolved, so this situation is not anomalous. We compared the properties of the component stars to predictions from {\sc parsec} models \cite{Bressan+12mn}, via the mass--radius and mass--\Teff\ diagrams \cite{MeClausen07aa,Me22obs4}. Assuming a solar chemical composition (fractional metal abundance $Z=0.017$), we found a decent match for an age of 1.7~Gyr. The \Teff\ values of the stars are slightly too low for their radii, and a better agreement would be obtained for \Teff\ values larger by 50~K. This is in line with the higher \Teff\ values found by Kang et al.\ \cite{Kang+12aj}. We plot a Hertzsprung-Russell diagram in Fig.~\ref{fig:hrd} which shows that the stars are reasonably consistent with the {\sc parsec} models and that they are evolved into the upper half of the main-sequence band.

%%%%%%%%%%%%%%%%%%%%%%%%%%%%%%%%%%%%%%%%%%%%%%%%%%%%%%%%%%%%%%%%%%%%%%%%%%%%%%%%%%%%%%%%%%%%%%%%%%%%%%%%%%%%%%%%%%%%%%%%%%%%%%%%%%%%%%%%%%%%%%%%%%%%%

\section*{Summary}

\targ\ is a well-known and extensively studied dEB containing two F3~V stars of very similar mass but significantly different radii orbiting with a period of 4.992~d. We have analysed the light curves of this object from two sectors of the TESS\ mission, obtaining photometric parameters to very high precision. We have reanalysed two available sets of RVs to determine the spectroscopic orbits of the stars and thus their full physical properties. Divergent \Teff\ determinations exist in the literature so we obtained our own by requiring the distance to the system to match that measured from its \gaia\ parallax.

Following standard conventions, we defined the primary eclipse to be deeper than the secondary eclipse, and star~A to be at inferior conjunction during primary eclipse. Although the eclipse depths are very similar they are measurably different in the TESS\ data so, for the first time, it is possible to unambiguously define which star is star~A. We find that it is hotter but smaller and less massive than its companion: the surface brightness ratio is convincingly below 1.0 whereas the mass ratio is conclusively above 1.0. We assembled a set of published and new times of minimum light and obtained a new ephemeris from which orbital phases can be calculated to high precision for the forseeable future.

The two stars have evolved into the second half of the main sequence band, and the greater evolution of star~B is clear. The properties of the system are consistent with the {\sc parsec} models for an age of 1.7~Gyr and a solar chemical composition. The similarity of the two stars, coupled with their slightly different evolutionary status, means \targ\ may be useful in future for helping to constrain and calibrate theoretical models of stellar evolution. A spectral analysis based on multiple high-quality spectra would be helpful to determine the atmospheric parameters of the two stars more accurately.

%%%%%%%%%%%%%%%%%%%%%%%%%%%%%%%%%%%%%%%%%%%%%%%%%%%%%%%%%%%%%%%%%%%%%%%%%%%%%%%%%%%%%%%%%%%%%%%%%%%%%%%%%%%%%%%%%%%%%%%%%%%%%%%%%%%%%%%%%%%%%%%%%%%%%

\section*{Acknowledgements}

We thank Drs.\ Guillermo Torres and Frank Fekel for help in our attempts to track down the RVs of \targ\ from Claud Lacy, and Zac Jennings for taking part in the INT/IDS observations.
This paper includes data collected by the TESS\ mission and obtained from the MAST data archive at the Space Telescope Science Institute (STScI). Funding for the TESS\ mission is provided by the NASA's Science Mission Directorate. STScI is operated by the Association of Universities for Research in Astronomy, Inc., under NASA contract NAS 5–26555.
 The following resources were used in the course of this work: the NASA Astrophysics Data System; the SIMBAD database operated at CDS, Strasbourg, France; and the ar$\chi$iv scientific paper preprint service operated by Cornell University.

%%%%%%%%%%%%%%%%%%%%%%%%%%%%%%%%%%%%%%%%%%%%%%%%%%%%%%%%%%%%%%%%%%%%%%%%%%%%%%%%%%%%%%%%%%%%%%%%%%%%%%%%%%%%%%%%%%%%%%%%%%%%%%%%%%%%%%%%%%%%%%%%%%%%%

\bibliographystyle{obsmaga}
% \bibliography{jkt}

%%%%%%%%%%%%%%%%%%%%%%%%%%%%%%%%%%%%%%%%%%%%%%%%%%%%%%%%%%%%%%%%%%%%%%%%%%%%%%%%%%%%%%%%%%%%%%%%%%%%%%%%%%%%%%%%%%%%%%%%%%%%%%%%%%%%%%%%%%%%%%%%%%%%%
\end{document}